\documentclass[aps,prb,showpacs,twocolumn,amssymb]{revtex4}

\usepackage{amsmath}
\usepackage{subfigure}
\usepackage{graphicx}% Include figure files

\input epsf

\def\prb{Phys. Rev. B }
\def\prl{Phys. Rev. Lett. }

\def\be{\begin{equation}}
\def\ee{\end{equation}}
\def\ba{\begin{eqnarray}}
\def\ea{\end{eqnarray}}
\def\ie{ {\it i.e.} }

\def\LSCO{La$_{2-x}$Sr$_x$CuO$_4$ }

\def\YBCO{YBa$_2$Cu$_3$O$_{7-\delta}$ }
\def\124{YBa$_2$Cu$_4$O$_8$ }

\def\BSCCO{Bi$_2$Sr$_2$CaCu$_2$O$_{8+\delta}$ }
\def\C60{A$_x$C$_{60}$ }

\def\LBCO{La$_{2-x}$Ba$_{x}$CuO$_4$ }

\begin{document}

\title{Spectral signatures of modulated $d$-wave superconducting phases}

\author{Shirit~Baruch and Dror~Orgad}

\affiliation{Racah Institute of Physics, The Hebrew University, Jerusalem 91904, Israel}

\date{\today}

\begin{abstract}

We calculate within a mean-field theory the spectral signatures of various striped
$d$-wave superconducting phases. We consider both in-phase and anti-phase modulations
of the superconducting order across a stripe boundary, and the effects of coexisting
inhomogeneous orders, including spin stripes, charge stripes, and modulated
$d$-density-wave. We find that the anti-phase modulated $d$-wave superconductor exhibits
zero-energy spectral weight, primarily along extended arcs in momentum space.
Concomitantly, a Fermi surface appears and typically includes both open segments and closed
pockets. When weak homogeneous superconductivity is also present the Fermi surface collapses
onto nodal points. Among them are the nodal points of the homogeneous $d$-wave superconductor,
but others typically exist at positions which depend on the details of the modulation and the
band structure. Upon increasing the amplitude of the constant component these additional
points move towards the edges of the reduced Brillouin zone where they eventually disappear.
The above signatures are also manifested in the density of states of the clean, and the
disordered system. While the presence of coexisting orders changes some details of the
spectral function, we find that the evolution of the Fermi-surface and the
distribution of the low-energy spectral weight are largely unaffected by them.

\end{abstract}


\maketitle

\section{Introduction}
\label{intro}

Over the last decade it has become increasingly clear that the cuprate
high-temperature superconductors exhibit inhomogeneous electronic structures
both in their "normal" and superconducting states \cite{ourreview}. In particular,
several scanning tunneling spectroscopy experiments have produced evidence for real-space
inhomogeneity of the superconducting gap without apparent spatial ordering
\cite{pan,lang,kato,mcelroy3,jamei,fang}, while others have demonstrated
periodic modulations of the local density of states (LDOS)
\cite{hoffman1,hoffman2,mcelroy1,howald,canacuocl,kap-ni,vershinin,mcelroy2,hashimoto,kohsaka}.
Even if much of the modulated signal can be attributed to interference patterns resulting
from scattering of quasiparticles off impurities \cite{hoffman2,mcelroy1,wang}, at least part
of it is likely associated with a spatially periodic structure of the superconducting
order and possibly of other coexisting electronic orders
\cite{howald,canacuocl,kap-ni,vershinin,mcelroy2,hashimoto,kohsaka,tesanovic}.

In this paper we study, within mean field theory, various realizations of
modulated $d$-wave superconducting (DSC) phases. We focus on stripe phases in which the
lattice-translational invariance is broken in one direction, and calculate their spectral
signatures in the clean limit and in the presence of disorder. The quasiparticle
density of states of clean translational-symmetry breaking states has been calculated
by Podolsky {\it et al.} in Ref. \onlinecite{podolsky}. There, the effects of a small modulated
order parameter added to a large-amplitude $d$-wave superconductor, were investigated. Here we
extend this treatment to strong modulations, consider also the distribution of spectral weight
in momentum space, and study additional scenarios which were not treated in
Ref. \onlinecite{podolsky}. In particular, we analyze the case of an anti-phase modulated
superconductivity, where the $d$-wave order parameter suffers a $\pi$-phase shift across each
stripe boundary. Such a state has been proposed recently \cite{himeda02,erez07} as the source for the
apparent decoupling between the Cu-O planes in 1/8 doped \LBCO \cite{kbt-lbco}.
Similar structures were also found in a mean-field study of the DSC
resonating valence bond phase of the $t-J$ model \cite{Raczkowski}. Finally, we supplement the
extensive literature on the Fourier-transformed local density of states (FT-LDOS)
of disordered superconducting phases, with and without coexisting inhomogeneous orders
\cite{bal-rev,tami,chen1,yichen1,bena,misra,check,atkinson,cheng,dellanna,goshal,yichen2,nunner},
by calculating the FT-LDOS of a system with anti-phase modulated superconductivity.

We demonstrate that in sharp contrast to its homogeneous and in-phase modulated counterparts,
the anti-phase modulated $d$-wave superconductor exhibits a Fermi surface which includes
extended parts of the non-interacting Fermi-surface, as well as closed pockets. The
corresponding zero-energy spectral weight appears predominantly over arcs in momentum space,
whose extent shrinks with increasing amplitude of the order parameter. We find that this
general behavior is robust with respect to changes in the details of the band-structure and
in the functional form of the modulations. Furthermore, the general evolution of the
Fermi-surface and the distribution of low-energy spectral weight is largely unaffected
by the introduction of additional coexisting orders. The situation changes significantly
when a homogeneous superconducting component is also present in the system. Under such
conditions the Fermi-surface collapses onto nodal points, which include those of the
homogeneous $d$-wave superconductor, and generically, additional points at positions
which depend on details of the modulation and band-structure. Upon increasing the constant
order parameter the extra nodal points move towards the edges of the reduce Brillouin zone (BZ),
where they eventually disappear. The presence of the Fermi-surface is also reflected in the
low-energy density of states of the anti-phase modulated DSC state, and in the momentum-space
structure of the FT-LDOS of this system when impurity scattering is included in the analysis.

\section{The Models}
\label{models}

Our starting point is a tight binding model of electrons hopping on a square lattice.
In the following we consider two non-interacting band structures, corresponding to
\BSCCO and \LBCO$\!\!$. While most of the scanning tunneling experiments have been carried
out on the first, owing to the good quality of surfaces which can be obtained in this
system, the latter is probably the most promising compound for the observation of the
signatures we discuss in the following, especially in the context of anti-phase modulated
superconductivity. Consequently, most of our results will be presented for this material.
To model \BSCCO we use the tight-binding Hamiltonian provided by Norman
{\it et. al.} \cite{norman}, for the system near optimal doping.
The 1/8 hole-doped \LBCO is described by the free dispersion
\cite{OK} $\xi(\mathbf{k})=-(t_1 /2) [\cos(k_x)+\cos(k_y)]-t_2\cos(k_x)\cos(k_y)-\mu$,
where $t_1=1.72$ eV, $t_2=-0.15t_1$ and $\mu$ is the chemical potential which we adjust
in order to maintain the required level of hole-doping.

In this work we study, using a mean-field Hamiltonian, the spectral signatures of a
spatially modulated $d$-wave superconductor. We also consider within the same approximation,
the effects of other inhomogeneous coexisting orders on such properties. Motivated by
the experimental evidence for the existence of "stripes" in the high-temperature
superconductors \cite{ourreview}, we analyze the case where the discrete translation
symmetry of the underlying lattice is broken by the periodic orders in the $x$ direction.
Since often the observed charge and spin modulation periods are 4 and 8 lattice constants,
respectively, we assume a unit cell of either lengths for the modulated orders. In the
following we describe the real-space structure of the various configurations which are
studied in the paper.

\subsection{The superconducting order}
\label{SC_models}

\begin{figure}[ht]
\includegraphics[angle=0,width=3.25in]{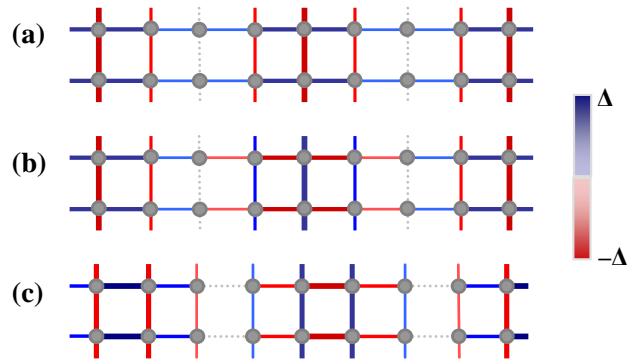}
\caption{The DSC configurations considered in this work: a) site-centered in-phase modulated,
and b) site-centered anti-phase modulated. c) bond-centered anti-phase modulated.
The magnitude of the order parameter is depicted by the width of the lines, while
its sign is given by the colors of the bonds.}
\label{DSCconfig}
\end{figure}

We consider two configurations for the DSC order. In the first its magnitude is sinusoidally
modulated across each site-centered stripe, but its phase is constant over the sample. This
"in-phase" configuration is described by the mean-field Hamiltonian
\ba
\nonumber
\!\!\!\!\!H_{0-d{\rm SC}}&=&\sum_{x,y}{\Bigg \{}\frac{\Delta}{4}
\left|\cos[q_x(x+1/2)]\right| \\
\nonumber
&\times&\left[ c_{x,y\uparrow}^\dag
c_{x+1,y\downarrow}^\dag - c_{x,y\downarrow}^\dag c_{x+1,y\uparrow}^\dag \right] \\
\nonumber
&-& \frac{\Delta}{4} \left|\cos(q_x x)\right| \\
&\times&\left[ c_{x,y\uparrow}^\dag c_{x,y+1\downarrow}^\dag
- c_{x,y\downarrow}^\dag c_{x,y+1\uparrow}^\dag \right]+{\rm H.c.}{\Bigg \}},
\label{in_phase_H}
\ea
where $x$ and $y$ are measured in units of the lattice constant which
we take from now on to be 1. Here, and throughout the paper $q_x=\pi/4$.
Note, that since the order parameter does not change its sign across the stripe boundary
the resulting configuration has a 4-site unit cell, as presented in Fig. \ref{DSCconfig}.
On a microscopic level it may reflect a variation in the pairing amplitude as a result of
a corresponding modulation of the hole density in the stripe phase, as discussed below.

The second, "anti-phase", configuration is similar to the first, except that the phase
of the order parameter changes by $\pi$ across a stripe boundary. It has been suggested in
Ref. \onlinecite{erez07}, that despite being unconventional, such a negative Josephson coupling
between stripes may occur in 1/8 doped \LBCO$\!$, with significant consequences for the
transport through this system. We consider the site-centered version of the
configuration, see Fig. \ref{DSCconfig}, and analyze the Hamiltonian
\ba
\nonumber
\!\!\!\!\!H_{\pi-d{\rm SC}}&=&\sum_{x,y}{\Bigg \{}\frac{\Delta}{4} \cos[q_x(x+1/2)] \\
\nonumber
&\times&\left[c_{x,y\uparrow}^\dag
c_{x+1,y\downarrow}^\dag -c_{x,y\downarrow}^\dag c_{x+1,y\uparrow}^\dag \right] \\
\nonumber
&-&\frac{\Delta}{4} \cos(q_x x) \\
&\times&\left[ c_{x,y\uparrow}^\dag c_{x,y+1\downarrow}^\dag -
c_{x,y\downarrow}^\dag c_{x,y+1\uparrow}^\dag \right]+{\rm H.c.}{\Bigg \}}.
\label{anti_phase_H}
\ea
The anti-phase nature of the configuration results in an 8-site unit cell.

We have also considered the bond-centered analogs of the above site-centered configurations.
Fig. \ref{DSCconfig} depicts, for example, the bond-centered anti-phase modulated DSC. We
have found, however, that the results are hardly affected by this detail, and therefore
in the following we report them for the site-centered orders.

\subsection{Coexisting electronic orders}

Various electronic orders have been proposed to exist in the cuprate high-temperature
superconductors, especially in the underdoped regime. In this paper we consider the case
where the modulated superconducting order is accompanied by one or more of the following:
charge stripes, anti-phase spin stripes and an anti-phase modulated $d$-density wave (DDW).

\begin{figure}[ht]
\includegraphics[angle=0,width=3.0in]{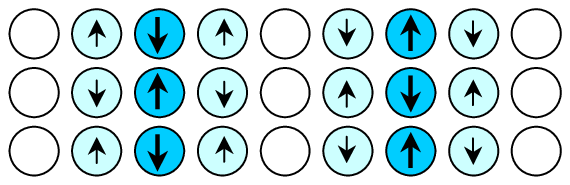}
\caption{A sketch of the magnetic and charge distribution in the stripe model. The arrows
represent the amplitude and sign of the spin density, and the shading of the circles
represents the charge density (the darker the circle is, the higher the charge density).}
\label{stripe}
\end{figure}

For the charge stripe phase we assume a site-centered, period-4 sinusoidal charge density
$\langle n_{x,y\uparrow} + n_{x,y\downarrow}\rangle = \phi_{CDW}\left|\sin(q_x x)\right|$.
When spin stripe order is also present the boundaries of the charge unit-cells serve
as anti-phase domain walls for an antiferromagnetic (AF) spin order of the form
$\langle n_{x,y\uparrow} - n_{x,y\downarrow}\rangle = \phi_{AF}(-1)^{x+y}\sin(q_xx)$.
The combined charge-spin stripe configuration is depicted in Fig. \ref{stripe}, and the
corresponding Hamiltonian is described in appendix \ref{app:models}.

\begin{figure}[ht]
\includegraphics[angle=0,width=3.0in]{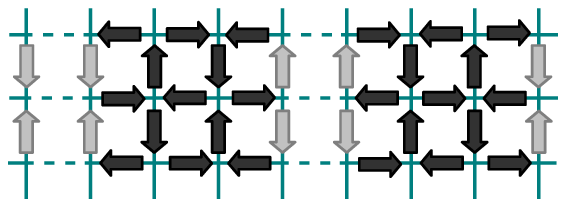}
\caption{A schematic representation of the modulated DDW order parameter.
The arrows represent the currents in the system. Gray arrows carry half the
current of the black ones.}
\label{DDW}
\end{figure}

The uniform DDW state was proposed as a model for the pseudogap
state in the underdoped cuprates\cite{DDW1}. It describes a
condensation of electron-hole pairs with non-zero angular momentum
into a phase with staggered current loops. In the following we
consider the anti-phase, period-8, modulated version of this state
as displayed in Fig. \ref{DDW}. Note that current conservation has
been incorporated into the model. For a detailed expressions of the
order parameter and the mean field Hamiltonian see appendix \ref{app:models}.

\section{Results}
\label{results}

\subsection{The Fermi-surface and the distribution of low-energy spectral weight}

\begin{figure}[ht]
{\includegraphics[width=3.4in]{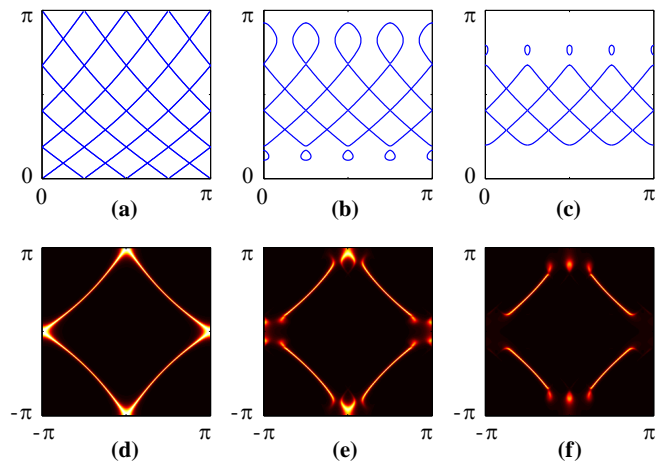}}
\caption{The Fermi Surface (a)-(c), and the low-energy spectral weight (d)-(f)
of a model corresponding to 1/8 hole-doped \LBCO in the presence of an anti-phase modulated
DSC. The Fermi surface is plotted in the first quadrant of the Brillouin zone. The spectral
weight is integrated within a 20 meV Lornetzian window centered at zero energy. The amplitude
of the modulated DSC is $\Delta=0$ in (a,d), $\Delta=0.075 t_1$ in (b,e) and
$\Delta=0.15 t_1$ in (c,f).}
\label{LBCO:FS} % caption for the whole figure
\end{figure}

\begin{figure}[ht]
{\includegraphics[width=3.4in]{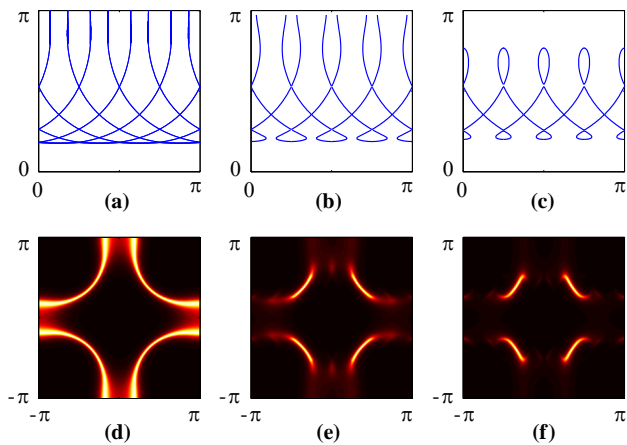}}
\caption{The Fermi Surface and the low-energy spectral weight in a model of 17\% hole-doped
\BSCCO with anti-phase modulated DSC. The spectral weight was integrated in a similar manner
to Fig. \ref{LBCO:FS}. The results are given for $\Delta=0$ in (a,d),
$\Delta=0.075 t$ in (b,e), and $\Delta=0.15 t$ in (c,f) ($|t|\simeq 0.6$ eV, see
Ref. \onlinecite{norman}.)}
\label{BSCCO:FS} % caption for the whole figure
\end{figure}

It is well known that the Fermi-surface of a homogeneous $d$-wave superconductor consists
of four nodal points. A very different behavior emerges in the case of an anti-phase
modulated DSC, as shown by Fig. \ref{LBCO:FS}. The figure exhibits the evolution of the
Fermi-surface of \LBCO with increasing strength, $\Delta$, of the anti-phase modulated DSC,
where the chemical potential was adjusted to maintain a 1/8 hole-doping level for each set of
parameters. One can clearly see that the model supports an extended Fermi-surface. As
the amplitude $\Delta$ increases, segments of the non-interacting
Fermi-surface near $(0,\pi)$ and symmetry related regions, become progressively gapped.
At the same time the Fermi-surface develops closed pockets which shrink with increasing
$\Delta$ and eventually disappear completely. The process proceeds until a single pocket
remains when $\Delta$ is of the order of the band-width. It then continues to shrink upon
further increase of the DSC order. This behavior is generic as can be seen in Fig.
\ref{BSCCO:FS}, where similar results are presented for \BSCCO. The rate at which the
Fermi-surface evolves, depends, however, on the band-structure. The flatter sections in
the anti-nodal regions of \BSCCO tend to develop a gap more quickly than the rest of the
Fermi-surface.

We believe that the origin of the Fermi-surface can be traced to the formation of
zero-energy Andreev bound states on the anti-phase domain walls of the DSC order.
A single junction between two phase-biased $d$-wave superconductors has been studied
by Tanaka and Kashiwaya\cite{tanaka1}. They have calculated the quasiparticle LDOS at the
interface between the superconductors and demonstrated that for a $\pi$-phase shift
the LDOS displays a pronounced zero-bias peak, associated with zero-energy Andreev
bound states. The latter move to higher energies when the phase difference is tuned
away from $\pi$, and correspondingly the LDOS exhibits a pseudo-gap behavior at low
biases. In our model we consider a chain of such junctions and therefore must take
into account the multiple scattering processes between them. We have not carried out
a detailed analytical analysis of this problem. Nevertheless, the numerical real-space
structure of the zero-energy wavefunctions reveal that the states residing near the
end-points of the open Fermi-surface segments are indeed localized on the domain walls.
On the other hand, the zero-energy states on the Fermi-pockets and away from these end
points are approximately evenly distributed over the system.
%While we have not carried
%out a detailed analysis of this problem we take the numerical results shown in Figs.
%\ref{LBCO:FS}, \ref{BSCCO:FS}, as a strong indication that the zero-energy Andreev
%states exist also in the multi-domain system.

Angle resolved photoemission spectroscopy (ARPES) is arguably the most suitable probe
for detecting the extended Fermi-surface. More generally, ARPES measures the
spectral function $A(\mathbf{k},\omega)$ of the system. In Figs. \ref{LBCO:FS}, \ref{BSCCO:FS}
we plot the distribution of low-energy spectral weight of an anti-phase modulated $d$-wave
superconductor, namely the frequency integrated $A(\mathbf{k},\omega)$ within
a 20 meV Lornetzian window centered at zero energy. As shown in the figures, the
spectral weight lies predominantly along arcs in momentum space. As we increase the
superconducting order the arcs shrink towards the BZ diagonals. For the
sinusoidal modulation studied here the distribution of spectral weight along the arcs
is continuous. However, for a square-wave modulation, gaps open on the arcs at the
boundaries of the reduced BZ at $k_x=\pm\pi/8$ and at points which are
related to them by the modulation wavevector.

\begin{figure}[th]
\includegraphics[angle=0,width=3.3in]{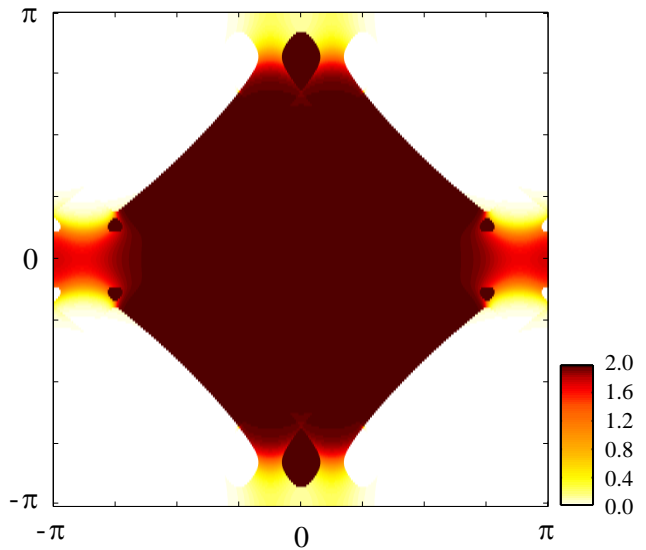}
\caption{The momentum occupation function $n(\mathbf{k})$ of a model of 1/8 hole-doped
\LBCO with an anti-phase modulated DSC of strength $\Delta=0.075t_1$.} \label{filling}
\end{figure}

\begin{figure}[th]
{\includegraphics[width=3.34in]{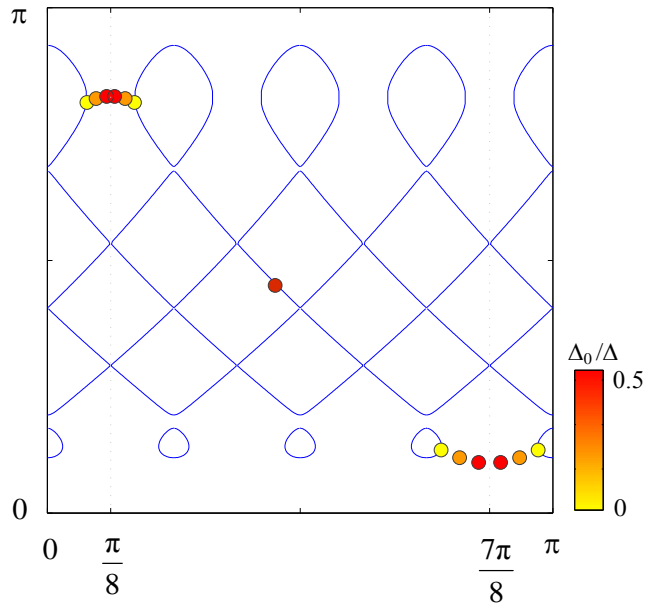}}
\caption{The evolution of the nodal points in a model of 1/8 hole-doped \LBCO with anti-phase
modulated DSC of strength $\Delta=0.075t_1$, as function of the amplitude, $\Delta_0$,
of an additional homogeneous DSC component. The curve depicts the Fermi-surface for
the case $\Delta_0=0$.}
\label{LBCO:nodal} % caption for the whole figure
\end{figure}

\begin{figure*}[th]
\includegraphics[width=6.4in]{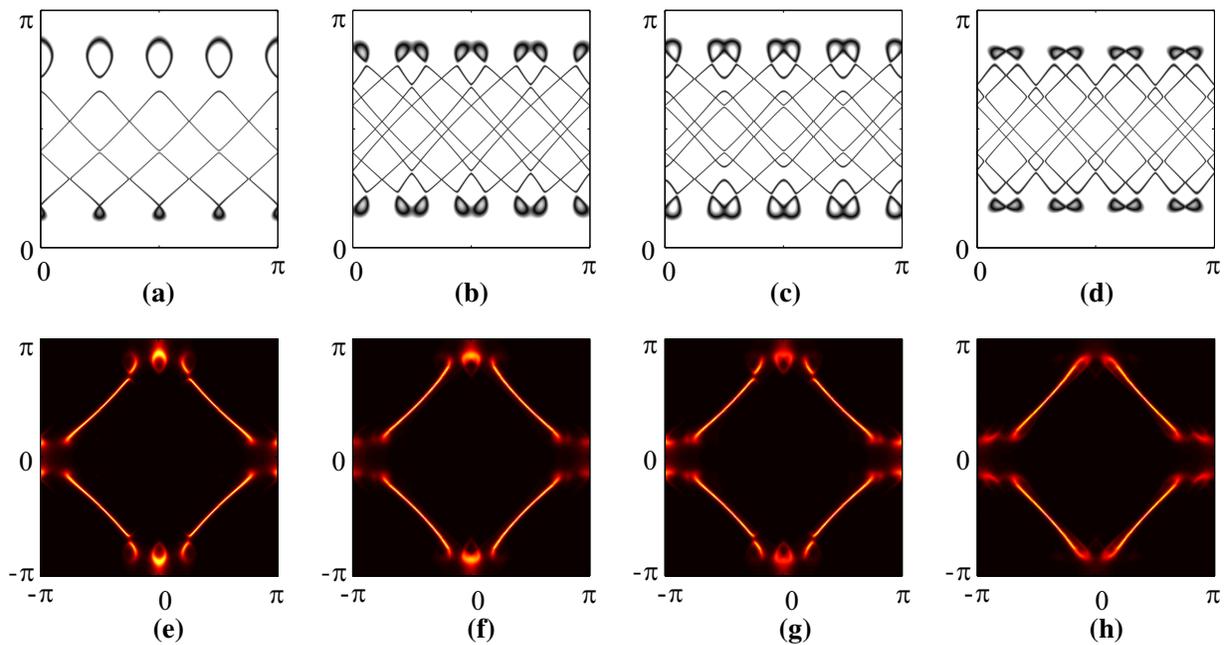}
\caption{Fermi surfaces (a)-(d) and low-energy spectral weight distributions (e)-(h) of a
model corresponding to 1/8 hole-doped \LBCO with anti-phase modulated DSC
and additional orders. The amplitude of the DSC order parameter is $\Delta=0.1 t_1$, and the
additional orders are: in (a,e) CDW with $\phi_{CDW}=0.075 t_1$, (b,f) spin-stripe order with
$\phi_{AF}=0.075 t_1$, (c,g) charge and spin stripe order with
$\phi_{AF}=\phi_{CDW}=0.075t_1$, (d,h) modulated DDW with $\phi_{DDW}=0.0125t_1$. Note that
the reduced BZ is halved in the presence of spin stripes or a modulated DDW, thus
complicating a straightforward comparison between the different Fermi surfaces.}
\label{extra orders} % caption for the whole figure
\end{figure*}

ARPES can also provide a measurement of the momentum occupation function
$n(\mathbf{k})=\int_{-\infty}^0 (d\omega/2\pi) A(\mathbf{k},\omega)$. It is this function
which depicts most directly the presence of the closed segments, or "pockets", in the
Fermi-surface. Fig. \ref{filling} displays $n(\mathbf{k})$ for the anti-phase modulated
DSC in \LBCO. While Fermi-surface pockets in the normal state can always be
classified as electron-like or hole-like, this is no longer true for the superconductor
whose quasiparticles are linear combinations of both electrons and holes. However, as
can be seen from Fig. \ref{filling}, the pockets at $k_x\simeq 0$ and $k_x\simeq \pm 3\pi/4$
are close to being electron-like, while the pockets at $k_x\simeq \pm \pi$ are of the hole-like
type. Recently, much interest has been generated by the observation of Shubnikov-de Haas and de
Haas-van Alphen oscillations in underdoped \YBCO in strong magnetic fields which suppress
superconductivity \cite{Doiron-lenard,Jaudet}. The experimental results suggest that the
Fermi-surface contains coherent electron-like pockets. Theoretical studies of non-superconducting
spin-charge modulated phases have found evidence for electron \cite{Millis_Norman} and hole \cite{Matz}
pockets.
Although unlikely to be relevant to the understanding of the aforementioned experiments,
it would be interesting to investigate the impact of the pockets which we find in the anti-phase
model, on the transport and magnetic properties of the system. We, however, did not pursue
this direction.
%Recently, electron\cite{Millis_Norman} and hole\cite{Matz} pockets were
%studied in theoretical non-superconducting models exhibiting spin-charge modulations.

When a constant DSC component, of arbitrary small magnitude, is added to the anti-phase
modulated system the Fermi-surface changes dramatically and collapses onto nodal points.
Among them we always find the nodal points of the uniform $d$-wave superconductor along
the diagonals of the BZ. However, as demonstrated by Fig. \ref{LBCO:nodal},
additional nodal points typically appear for not too large magnitude, $\Delta_0$, of the
constant component. For the parameters used by us their spectral weight is about a half
of the weight of the points along the diagonals. When $\Delta_0$ is increased, the additional
points move from the tips of the pockets, where they originate, towards the edge of the
reduced BZ where they disappear, while the points along the diagonals stay put. The number
and positions of the nodal points vary as function of the strength of the modulated order
and of the band-structure, but the general behavior remains the same. In the case of an
in-phase modulated DSC we always find a Fermi-surface which consists of the four nodal
points of the uniform superconductor. This is not surprising since the in-phase modulated
order can be represented approximately as an equal strength combination of anti-phase and
constant components, for which, as seen in Fig. \ref{LBCO:nodal}, the extra nodal points have
long vanished. It is worth mentioning that while the change in the topology of the Fermi-surface
takes place even for an infinitesimal $\Delta_0$, the original Fermi-surface of the anti-phase
modulated system is only gapped by an amount of order $\Delta_0$. Therefore, the low-energy
spectral weight distribution of a system with weak additional homogeneous DSC is close to
the one presented in Figs. \ref{LBCO:FS}, \ref{BSCCO:FS}.

We have also studied the Fermi-surface and the low-energy spectral weight in the presence
of additional electronic orders, as described in section \ref{models}. At large, we have
found that the coexisting orders do not change the general features outlined above. The
Fermi-surface of the in-phase modulated system is still comprised of the conventional
$d-$wave nodal points. The introduction of the orders into the anti-phase modulated
superconductor does have some effect on the Fermi-surface and the spectral weight
distribution, as seen in Fig. \ref{extra orders} (Note that the reduced BZ is halved in the
presence of spin stripes or a modulated DDW.) Most notably, we observe that the CDW order
induce gaps on the Fermi-surface at multiples of the ordering wavevector. Although the shape
of the Fermi-surface is somewhat different, its extent is largely unaffected by the additional
orders. This is also true for the general evolution of the Fermi-surface with increasing
$\Delta$ and with the addition of a uniform DSC component.

\subsection{The density of states}

\begin{figure}[b]
{\includegraphics[width=3.2in]{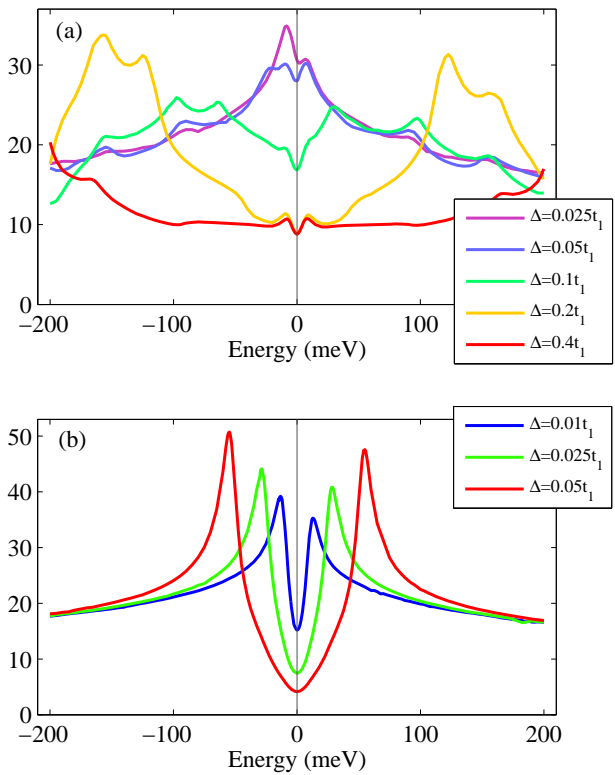}}
\caption{Averaged density of states in a model of LBCO with (a) anti-phase modulated DSC and
(b) in-phase modulated DSC, for various magnitudes of the superconducting order.}
\label{REAL_LDOS} % caption for the whole figure
\end{figure}

\begin{figure}[ht]
\includegraphics[angle=0,width=3.2in]{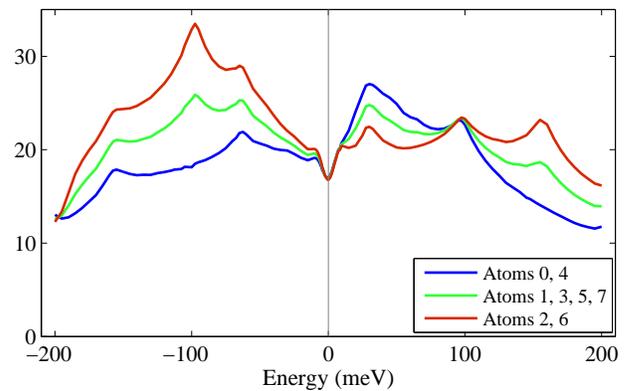}
\caption{LDOS in a model of \LBCO with an anti-phase modulated DSC of strength $\Delta=0.1t_1$.
Atom 0 is located on the symmetry axis of the stripe - the left most site in Fig.
\ref{DSCconfig}.}
\label{LDOS}
\end{figure}

\begin{figure}[ht]
{\includegraphics[width=3.2in]{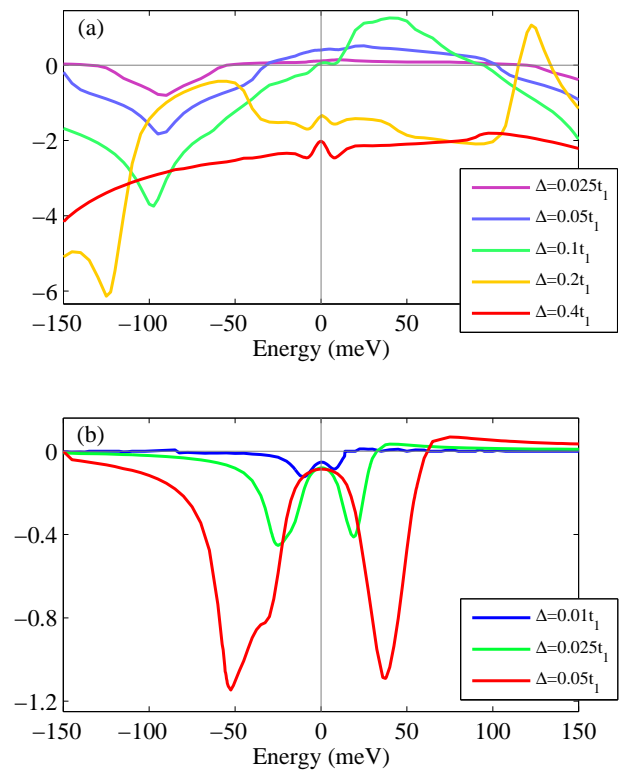}}
\caption{$\rho_{(\pi/2,0)}$ in a model of LBCO with (a) anti-phase modulated DSC and
(b) in-phase modulated DSC, for various magnitudes of the superconducting order.}
\label{half_pi} % caption for the whole figure
\end{figure}

The low-energy signatures discussed above are also manifested in the LDOS, which can be measured
using scanning tunneling spectroscopy. Fig. \ref{REAL_LDOS} presents the density of states of
the modulated superconducting states after averaging over the position in the sample. The
energy dependence of the DOS in the case of the in-phase modulated DSC is close to that
of a uniform $d$-wave superconductor whose amplitude is given by the zero-wavevector component
of the modulated order. In particular, the DOS vanishes at zero-bias (In order to simulate the
finite energy resolution of the experiments we have introduced a 5 meV Lorentzian broadening
into our numerical calculation. This is the reason for the apparent zero-energy DOS in the
figure.) In contrast, the zero-energy DOS of the anti-phase modulated system remains finite.
We find that the low-energy behavior of the DOS depends on the band-structure. For \LBCO the
DOS contains a small dip at zero bias whose width does not scale with the
strength of the modulated superconducting order, see Fig. \ref{REAL_LDOS}. In \BSCCO
on the other hand (not shown here), there is a more pronounced dip which does scale with
$\Delta$. In both cases the low-energy DOS decreases when $\Delta$ is increased, in agreement
with the reduction in the extent of the Fermi-surface, as described above.

Although the unit-cell in the anti-phase superconducting system is composed of 8 sites,
the real space periodicity of the LDOS is halved, as shown in Fig. \ref{LDOS}. This is a
result of the symmetries of the model. Translating the system by 4 lattice constants is
equivalent to the change $\Delta\rightarrow-\Delta$, which does not affect the LDOS, as
long as the average pairing potential is zero. An additional constant DSC component
breaks this symmetry, opens a $d$-wave gap in the LDOS, and reverts its period to 8 sites.
Consequently, the first non-vanishing Fourier component of the LDOS, ($\rho_q(\omega)$),
in both the in-phase and the pure anti-phase modulated models is at $\mathbf{q}=(\pi/2,0)$.
Its energy dependence is shown in Fig. \ref{half_pi}. While for the in-phase model it exhibits
a symmetric structure around zero bias within the low-voltage regime, such symmetry is absent
in the anti-phase case.

The addition of electronic orders to the superconducting system does not change the
main features described above. Zero-bias spectral weight is still observed in the
anti-phase DSC model and is absent in the in-phase system. The periodicity of the
LDOS is also unaffected by the CDW and the DDW orders, while the spin-stripe order
introduces new Fourier components at multiples of $\mathbf{q}=(\pi/4,\pi)$.

\subsection{FT-LDOS}

In recent years, scanning tunneling spectroscopy measurements of the FT-LDOS have become
a much-used tool in the quest for identifying unconventional electronic orders in the cuprates
\cite{bal-rev}. We have calculated the effects of quasiparticle scattering off a single impurity
on the FT-LDOS in a system with modulated superconductivity. To this end we have implemented the
T-matrix approximation, as described in appendix \ref{app:sts}.

\begin{figure}[ht]
{\includegraphics[width=3.3in]{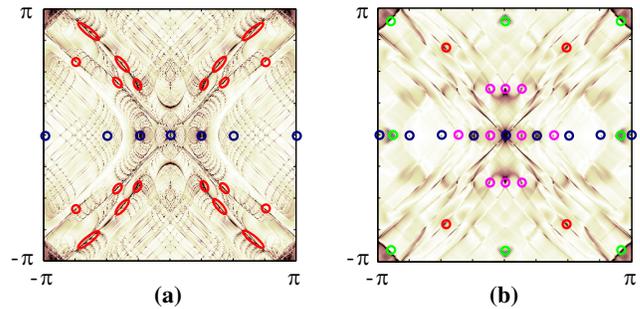}}
\caption{FT-LDOS at $\omega=0$ of a model of LBCO with (a) anti-phase modulated DSC order and
(b) coexisting anti-phase and homogeneous DSC orders. The amplitude of the modulated
order is $\Delta=0.1 t_1$ and that of the constant component is $\Delta_0=0.01 t_1$.
The results are for an impurity strength of $V_0=150$ meV.
Darker regions correspond to higher FT-LDOS and we have indicated high-intensity peaks
by circles: Peaks associated with the modulation wave vector are encircled in blue.
Peaks that correspond to scattering between the nodal points of the homogeneous DSC are
marked in green. Peaks due to scattering between the additional nodal points which appear
in the mixed phase are marked in pink. Other high-intensity peaks which are not simply
related to scattering between nodal points are designated in red.}
\label{STS-figure}
\end{figure}

Since most of the spectral changes in the modulated system occur at low energies we choose
here to concentrate on the zero-energy FT-LDOS, and present, in Fig. \ref{STS-figure}, its
distribution for an anti-phase modulated DSC system with, and without, an additional
homogeneous DSC component. As can be seen, the Fourier transformed spectra is highly detailed.
However, this fine structure is unlikely to be accessible experimentally as it is of very
small weight compared to the prominent peaks which dominate the distribution. Consequently,
we concentrate on the latter.

In both cases studied by us the strongest peaks appear at multiples of the modulation
wavevector $\mathbf{q}=(\pi/4,0)$. [Note that disorder eliminates the symmetry which produces the
period-4 LDOS of the clean anti-phase modulated system, thus enabling the appearance of the
peaks at $(\pm\pi/4,0)$.] The positions in $\mathbf{k}$-space of many of the remaining peaks
may be associated with wavevectors connecting high zero-energy LDOS regions of the clean
system \cite{hoffman2,wang}. This is especially true for the case with additional homogeneous
DSC where the zero-energy spectral weight appears at isolated nodal points. We were able
to identify, as shown in Fig. \ref{STS-figure}, peaks which are associated with scattering
between the nodal points of the homogeneous $d$-wave superconductors, as well as peaks which
originate from scattering between the extra nodal points that are induced by the presence of
the modulated DSC order, see Fig. \ref{LBCO:nodal}. It is more difficult to trace the root of
the peaks in the FT-LDOS of the pure modulated system, since it possesses an extended
Fermi-surface.

\section{Conclusions}
\label{conclusions}

We have identified several signatures which set apart the anti-phase modulated
$d$-wave superconductor from other $d$-wave superconducting states. Most notably,
we have shown that it supports a continuum of zero-energy excitations which form an
extended Fermi-surface. The existence of such a Fermi-surface requires a pure
$\pi$-phase shift of the superconducting order across the domains boundaries.
The presence of a constant DSC order, no matter how weak, drastically modifies
the manifold of zero-energy states and results in a Fermi surface which consists
of isolated nodal points. Nevertheless, much of the low-energy spectral weight still
concentrates along extended arcs in momentum space, and can be detected by ARPES measurements.
ARPES can also be used to identify the additional nodal points which we predict exist
under such circumstances, beside the conventional nodal points associated with a
homogeneous $d$-wave order. The periodic modulation of the superconducting order can be traced
using scanning tunneling spectroscopy, which can also detect the accompanying low-energy
spectral weight, and the identifying features in the FT-LDOS of the anti-phase modulated
superconductor.

\acknowledgments{It is a pleasure to thank Erez Berg, Cheng-Chien Chen and Steve Kivelson
for useful discussions and for sharing with us their results. This work was supported by
the United States - Israel Binational Science Foundation (grant No. 2004162).}

\appendix

\section{The Hamiltonians}
\label{app:models}

\subsection{Superconducting orders}

The analysis of the various orders was carried out using mean-field Hamiltonians written in
momentum space. The modulated superconducting orders are conveniently expressed in terms of the spinor
$\psi_\mathbf{k}^\dagger=(c_{\mathbf{k}\uparrow}^\dagger, c_{\mathbf{k}+\mathbf{q}\uparrow}^\dagger,
\ldots,c_{-\mathbf{k}\downarrow},c_{-(\mathbf{k}+\mathbf{q})\downarrow}, \ldots)$, with
$\mathbf{q}$ the ordering wavevector, and $\mathbf{k}$ varying over the reduced Brillouin zone (RBZ)
associated with the order. In this basis the Hamiltonian
\be
H=\sum_{\mathbf{k}\in RBZ} \;\psi_\mathbf{k}^\dagger \;\hat{H}_\mathbf{k} \;\psi_\mathbf{k},
\label{Hrep}
\ee
takes the form
\ba
\hat{H}_\mathbf{k}=\left( \begin{array}{cc}
\mathcal{A}_\mathbf{k} & \mathcal{C}_\mathbf{k} \\
   \mathcal{C}^\dagger_\mathbf{k} & -\mathcal{A}_\mathbf{k}
\end{array} \right),
\label{Hmatrix}
\ea
where $\mathcal{A}_\mathbf{k}$ is a diagonal matrix with entries $\xi(\mathbf{k}),
\xi(\mathbf{k}+\mathbf{q}),\ldots$, and where $\mathcal{C}_\mathbf{k}$ contains the
particle-particle couplings.

While the RBZ of the in-phase modulated superconductor is twice that of its anti-phase
counterpart, we choose to describe both in the RBZ of the latter, i.e., $-q_x/2<k_x<q_x/2$ and
$-\pi<k_y<\pi$, where $\mathbf{q}=(\pi/4,0)$. In this case $\mathcal{C}_\mathbf{k}$
is an $8\times 8$ matrix,
\ba
\!\!\mathcal{C}_\mathbf{k}=\left( \begin{array}{cccc}
f_0(\mathbf{k}) & f_1(\mathbf{k}) &\cdots & f_7(\mathbf{k}) \\
 f_1^*(\mathbf{k}) & f_0(\mathbf{k}+\mathbf{q}) & \cdots &  f_6(\mathbf{k}+\mathbf{q}) \\
  \vdots  &  &  \ddots   \\
 f_7^*(\mathbf{k}) &   &   &  f_0(\mathbf{k}+7\mathbf{q})
\end{array} \right)\!,
\label{Cmatrix}
\ea
characterized by the functions $f_n(\mathbf{k})$.

The only non-vanishing contribution to the uniform superconducting order is
\be
f_0(\mathbf{k})=\frac{\Delta_0}{2}(\cos k_x-\cos k_y).
\ee

The real-space representation of the in-phase modulated superconducting Hamiltonian is given
by Eq. (\ref{in_phase_H}). Since only integer $x$ and $y$ are of interest, when converting it to
the momentum-space representation, we approximate the absolute value function by its first eight
Fourier components, with the result
\ba
f_0(\mathbf{k})&=&\frac{\Delta}{8}
\left[2\sqrt{2}\cos(q_x/2)\cos k_x-(1+\sqrt{2})\cos k_y\right],\nonumber\\
f_2(\mathbf{k})&=&\frac{\Delta}{8}\left[\sqrt{2}\sin(q_x/2)(\cos k_x-\sin k_x)
-\cos k_y\right],\nonumber\\
f_4(\mathbf{k})&=&\frac{\Delta}{8}(-1+\sqrt{2})\cos k_y,\nonumber\\
f_6(\mathbf{k})&=&f_2(\mathbf{k}+6\mathbf{q}).
\ea

In the presence of an anti-phase modulated superconducting order one obtains
\ba
f_1(\mathbf{k})&=&\frac{\Delta}{4}\left[\cos (k_x+q_x/2)-\cos k_y\right],\nonumber\\
f_7(\mathbf{k})&=&\frac{\Delta}{4}\left[\cos (k_x-q_x/2)-\cos k_y\right].
\ea

\subsection{Non-superconducting orders}

For the Hamiltonians of the non-superconducting orders we use a similar representation to
Eq. (\ref{Hrep}), but with $\psi_\mathbf{k}$ defined with respect to the wavevector
$\tilde{\mathbf{q}}=(\pi/4,\pi)$. Here, we choose to let $\mathbf{k}$ run over the RBZ
of the modulated DDW and spin-stripe orders, despite it being half the size of the RBZ
of the charge-stripe phase. The result is
\ba
\hat{H}_\mathbf{k}=\left( \begin{array}{cc}
\mathcal{A}_\mathbf{k}+\mathcal{B}_\mathbf{k} &0 \\
   0 & -\mathcal{A}_\mathbf{k}-\eta\mathcal{B}_\mathbf{k}
\end{array} \right),
\label{Hmatrix-non}
\ea
where $\eta=1$ for the charge and DDW stripe phase, and $\eta=-1$ in the case of spin-stripe order.
$\mathcal{B}_\mathbf{k}$ contains the particle-hole couplings
\ba
\!\!\mathcal{B}_\mathbf{k}=\left( \begin{array}{cccc}
g_0(\mathbf{k}) & g_1(\mathbf{k}) &\cdots & g_7(\mathbf{k}) \\
 g_1^*(\mathbf{k}) & g_0(\mathbf{k}+\tilde{\mathbf{q}}) & \cdots &  g_6(\mathbf{k}+\tilde{\mathbf{q}}) \\
  \vdots  &  &  \ddots   \\
 g_7^*(\mathbf{k}) &   &   &  g_0(\mathbf{k}+7\tilde{\mathbf{q}})
\end{array} \right)\!,
\label{Bmatrix}
\ea

The non-vanishing contributions in the charge-stripe phase are
\ba
g_0(\mathbf{k})&=&\frac{\Phi_{CDW}}{4}\left[1+2\cos (\tilde{q}_x)\right],\nonumber\\
g_2(\mathbf{k})&=&g_6(\mathbf{k})=-\frac{\Phi_{CDW}}{4},\nonumber\\
g_4(\mathbf{k})&=&\frac{\Phi_{CDW}}{4}\left[1-2\cos (\tilde{q}_x)\right]\;,
\ea
where the absolute value function was again approximated using the low harmonics.
The $g_0(\mathbf{k})$ term renormalizes the chemical potential. The latter was chosen to
maintain the desired hole doping.

The spin-stripe phase is characterized by the following terms
\ba
g_3(\mathbf{k})=-g_5(\mathbf{k})=i\frac{\Phi_{AF}}{4}.
\ea

We considered an anti-phase, period-8, modulated version of the d-density wave,
(see Fig. \ref{DDW}). The corresponding real-space order parameter is
\ba
\nonumber
\langle c_{\mathbf{r},\sigma}^\dag c_{\mathbf{r}',\sigma}\rangle &=&\! i\frac{\Phi_{DDW}}{2}(-1)^{x+y} \\
&\times&\!\bigg\{2\delta_{\mathbf{r}',\mathbf{r}+\hat{x}}\Theta_{\tilde{q}_x}(x)
+2\delta_{\mathbf{r}',\mathbf{r}-\hat{x}} \Theta_{\tilde{q}_x}(x-1)\nonumber\\
&-&\!\left(\delta_{\mathbf{r}',\mathbf{r}+\hat{y}} + \delta_{\mathbf{r}',\mathbf{r}-\hat{y}}\right)
\left[\Theta_{\tilde{q}_x}(x)+\Theta_{\tilde{q}_x}(x-1)\right]\!\bigg\},\nonumber\\
\ea
where $\Theta_{q_x}(x)$ is a periodic variation of the usual heavyside function, \ie
\ba
\Theta_{q_x}(x)=\left\{
\begin{array}{cl}
1 & \;2\pi n/q_x< x< (2n+1)\pi/q_x\\
-1 & \;(2n-1)\pi/q_x< x< 2\pi n/q_x\\
0 & \;x=n\pi/q_x
\end{array}\right.
\ea
for all integers $n$. The corresponding functions entering the momentum-space Hamiltonian
are given by
\ba
g_1(\mathbf{k})&=& h(\mathbf{k},5\tilde{q}_x/2),\nonumber\\
g_3(\mathbf{k})&=& h(\mathbf{k},7\tilde{q}_x/2),\nonumber\\
g_5(\mathbf{k})&=& h(\mathbf{k},-7\tilde{q}_x/2),\nonumber\\
g_7(\mathbf{k})&=& h(\mathbf{k},-5\tilde{q}_x/2)\;,
\ea
where
\ba
h(\mathbf{k},q_x)&=&\Phi_{DDW}e^{i q_x}\sin(4q_x)[1 + 2\cos(2q_x)]\nonumber \\
&\times&[\cos(k_x + q_x)- \cos(k_y)\cos(q_x)].
\ea

\subsection{Combining the orders}

When superconducting and non-superconducting orders coexist we need to expand
our basis to include them both. We use
\ba
\psi_\mathbf{k}^\dagger&=&(c_{\mathbf{k}\uparrow}^\dagger, c_{\mathbf{k}+(q_x,0)\uparrow}^\dagger,
\ldots, c_{\mathbf{k}+(7q_x,0)\uparrow}^\dagger,c_{\mathbf{k}+(q_x,\pi)\uparrow}^\dagger,\ldots,
\nonumber\\
&&c_{\mathbf{k}+(7q_x,\pi)\uparrow}^\dagger, c_{-\mathbf{k}\downarrow}, c_{-\mathbf{k}+(q_x,0)\downarrow},
\ldots,c_{-\mathbf{k}+(7q_x,\pi)\downarrow}).\nonumber\\
\ea
The combined Hamiltonian is 32-dimensional and is defined in the momentum-space region $-q_x/2<k_x<q_x/2$,
and $-\pi/2<k_y<\pi/2$. The terms in this Hamiltonian are determined by adding the corresponding terms,
connecting the same momenta, in Hamiltonians (\ref{Hmatrix}) and (\ref{Hmatrix-non}).

\section{Calculating the FT-LDOS}
\label{app:sts}

The clean system
can be described in terms of the spinor $\psi_\mathbf{k}^\dagger$ defined in appendix
\ref{app:models}. In this basis the Hamiltonian is $H=\sum_{\mathbf{k}\in RBZ}
\psi_\mathbf{k}^\dagger \hat{H}_\mathbf{k} \psi_\mathbf{k}$,
where $\hat{H}_\mathbf{k}$ is a model dependent matrix. The Green's function of the clean system,
$\hat{G}^0(\mathbf{k},\omega)$,
is obtained by analytically continuing $\hat{G}^0(\mathbf{k},i\omega_n)
=[i\omega_nI-\hat{H}_\mathbf{k}]^{-1}$, via $i\omega_n\rightarrow\omega+i\delta$, where $I$
is the identity matrix. In the numerical calculations we have used an energy
broadening of $\delta=$0.5 meV.

The impurity scattering problem can be simplified by working in the Nambu basis in real space,
$(c^\dagger_{\mathbf{r}\uparrow}, c_{\mathbf{r}\downarrow})$. The transformation of the Green's
function between the two representations is given by
\ba
\label{green_transform}
&&\hat{G^0}(\mathbf{r},\mathbf{r}';\omega)=\frac{1}{(2\pi)^2}
\sum_{\mathbf{k}\in RBZ}e^{i\mathbf{k}(\mathbf{r}-\mathbf{r}')} \\
\nonumber
&&\times
\left( \begin{array}{cc}
\phi^\dagger(\mathbf{r})G^0_{11}(\mathbf{k},\omega)\phi(\mathbf{r}') & \phi^\dagger(\mathbf{r})
G^0_{12}(\mathbf{k},\omega)\phi(\mathbf{r}')\\
\phi^\dagger(\mathbf{r})G^0_{21}(\mathbf{k},\omega)\phi(\mathbf{r}') & \phi^\dagger(\mathbf{r})
G^0_{22}(\mathbf{k},\omega)\phi(\mathbf{r}')
\end{array}\right),
\ea
where $\phi^\dagger(\mathbf{r}) = (1,e^{i\mathbf{qr}},e^{2i\mathbf{qr}},\ldots)$,
and $G^0_{ij}$, $i,j=1,2$, are the four blocks comprising $\hat{G}^0(k,\omega)$.

We have included the impurity scattering in the T-matrix approximation \cite{bal-rev},
which implies the following equation for the electronic Green's function
\ba
\label{full_green}
&&\hat{G}(\mathbf{r},\mathbf{r}';\omega_n)=\hat{G}^0(\mathbf{r},\mathbf{r}';\omega) \\
\nonumber
&&+\int d\mathbf{r}_1 d\mathbf{r}_2 \hat{G}^0(\mathbf{r},\mathbf{r}_1;\omega)
\hat{T}(\mathbf{r}_1,\mathbf{r}_2;\omega)\hat{G}^0(\mathbf{r}_2,\mathbf{r}';\omega) .
\ea
For simplicity we have assumed scattering off a non-magnetic $\delta$-function impurity
centered at the origin. Under such conditions the T-matrix is given by
$\hat{T}(\mathbf{r},\mathbf{r}';\omega)=[V_0^{-1}\tau_3-\hat{G}^0(0;\omega)]^{-1}\delta(\mathbf{r})
\delta(\mathbf{r}')$,
where $V_0$ is the impurity strength and $\tau$ are the Pauli matrices.
The local density of states at a point $\mathbf{r}$ in the sample is then given by
$\rho(\mathbf{r},\omega)=-\frac{1}{\pi}{\rm Im}[G_{11}(\mathbf{r},\mathbf{r};\omega)+
G_{22}(\mathbf{r},\mathbf{r};-\omega)]$,
from which the FT-LDOS is readily obtained.

\end{document}